\documentclass{PoS}
\usepackage{citesort}

\title{The role of the running coupling constant in the unveiling of the hadronic structure.}

\ShortTitle{Running coupling constant and hadronic structure.}

\author{\speaker{A. Courtoy}\thanks{This work was started in Pavia (Italy), under the affiliation "INFN-Pavia".}\\
        IFPA, Inst. de Physique, Universit\'e de Li\`ege, Belgium \\
        E-mail: \email{aurore.courtoy@ulg.ac.be}}

\abstract{One of the main open questions in physics is the understanding of the internal structure of the strongly interacting particles, or hadrons. It is still a challenge to describe consistently the dynamics of scattering processes and hadronic structure at moderate energy scales.  The study of Parton Distribution Functions (PDFs) sets a connection
between the perturbative and non-perturbative worlds, through the following scheme: one builds models consistent with QCD in a moderate energy range, PDFs are evaluated in these models, and, finally, the scale dependence of these distributions is studied. In these proceedings, we revisit the standard procedure to match non-perturbative models to perturbative QCD, using experimental data. The strong coupling constant plays a central role in the QCD evolution of parton densities. We will extend this procedure with a non-perturbative generalization of the QCD running coupling and use this new development to understand why perturbative treatments are working reasonably well in the context of hadronic models. Vice versa, this new procedure broadens the ways of analyzing the freezing of the running coupling constant.}

\FullConference{International Workshop on QCD Green's Functions, Confinement and Phenomenology,\\
		September 05-09, 2011\\
		Trento Italy}

\begin{document}

    \section{Parton Distributions' Scales}
    
 The internal structure of the strongly interacting particles remains veiled. It is still a challenge to describe consistently the dynamics of scattering processes and hadronic structure at moderate energy scales. Because  at such a  scale a hadronic representation takes over the partonic description, it is called {\it the hadronic scale}. The hadronic scale is peculiar to each hadronic representation.
 
     A way of connecting the low and high-energy worlds has traditionally been through the study of Parton Distribution Functions (PDFs): 
    Deep Inelastic processes are such that they enable us  to look with a good resolution inside the hadron and allow us to resolve the very short distances, i.e. small configurations of quarks and gluons. At short distances, this part of the process is described through perturbative QCD.  A resolution of such short distances is obtained with the help of non-strongly interacting probes.  Such a probe, typically a photon, is provided by hard reactions. In that scheme, the PDFs reflect how the target reacts to the probe, or how the quarks and gluons are distributed inside the target. The insight into the structure of hadrons is reached at that stage:  the large virtuality of the photon, $Q^2$, involved in such processes allows for the factorization of  the hard (perturbative) and soft (non-perturbative) contributions in their amplitudes. Hence, the virtuality of the photon introduces another scale, i.e.,  {\it  the factorization scale}.

    The evaluation of PDFs  is guided by a standard scheme, set up in valuable literature of the 90s \cite{Traini:1997jz,Stratmann:1993aw,Parisi:1976fz}. This scheme runs in 3 main steps.
    First, we either build models consistent with QCD in a moderate energy range, typically the hadronic scale; or we use effective theories of QCD for the description of hadrons at  the same  energy range. Second,  PDFs are evaluated in these models, giving a description of the Bjorken-$x$ dependence of the distribution. Third, the scale dependence of these distributions is studied. The last step allows to bring the moderate energy description of hadrons to the factorization scale, thanks to the QCD evolution equations. In these proceedings, we will focus on this third step: how to match non-perturbative models to perturbative QCD, using experimental data.
    
    Besides, it is worth mentioning that, since the hadronic models are characterized  by their ingredients and degrees of freedom, they may also include the concepts of chiral symmetry breaking (e.g. NJL) or a description of confinement (e.g. MIT bag model). In either cases, those concepts are related to yet another momentum scale (respectively the {\it chiral symmetry breaking scale} and the {\it confinement scale}) that is peculiar to the model.

    
    \section{The Hadronic Scale}
    
 The hadronic scale is defined at a point where the partonic content of the model, defined through the second moment of the parton distribution,  is known. For instance, the CTEQ parameterization gives~\footnote{MSTW gives a similar result.}
 \begin{equation}
 \left \langle  (u_v+d_v)(Q^2=10 \mbox{GeV}^2)) \right\rangle_{n=2}=0.36\quad,
 \end{equation}
 with $q_v$ the valence quark distributions and with $\langle q_v (Q^2)\rangle_n=\int_0^1 dx \, x^{n-1} \, q_v(x, Q^2)$.
  In the extreme case,  i.e., when we assume that the partons are pure valence quarks, we evolve downward the second moment until
   \begin{equation}
 \left \langle  (u_v+d_v)(\mu_0^2) \right\rangle_{n=2}=1\quad.
 \end{equation}
 The hadronic scale is found to be $\mu_0^2 \sim 0.1$ GeV$^2$.  
   
   This standard procedure to fix the hadronic (non-perturbative) scale  pushes perturbative QCD to its limit. 
   In effect, the hadronic scale turns out to be of a few hundred MeV$^2$, where the strong coupling constant has already started approaching its Landau pole. As it will be shown hereafter, the N$^m$LO evolution converges very fast, what justifies the perturbative approach.
   Consequently, the behavior of the strong coupling constant plays a central role in the QCD evolution of parton densities.  Here we extend the standard procedure with the non-perturbative generalization of the QCD running coupling~\cite{Courtoy:2011mf}. 

\section{The Running Coupling Constant}

\begin{center}
\begin{figure}[tb]
\includegraphics[scale= .82]{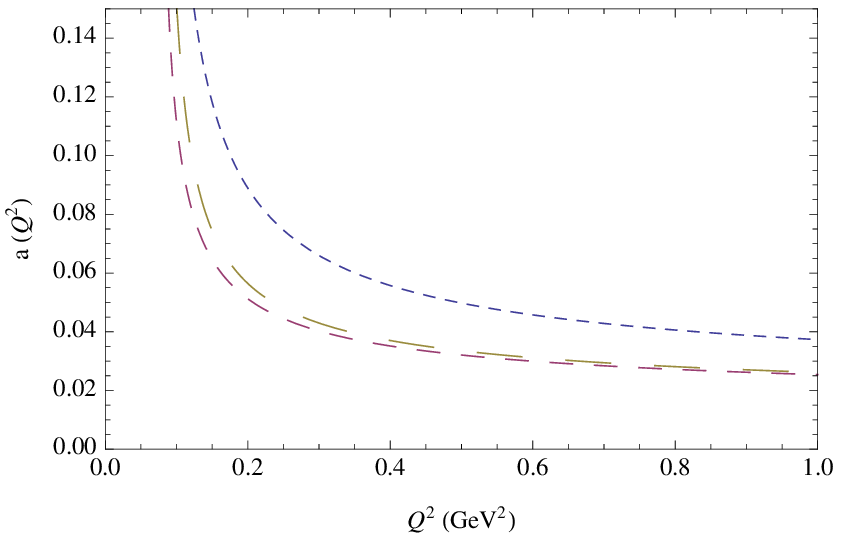}
\includegraphics[scale= .82]{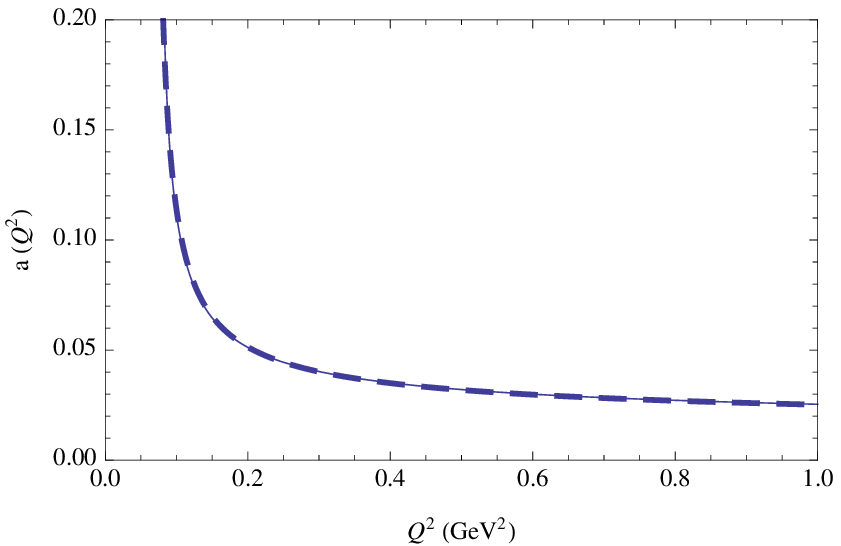}
\caption{ The running of the  coupling constant. Left: The short dashed curve  corresponds to the LO solution, the medium dashed curve to NLO solution and the long dashed curve to the NNLO solution($\Lambda = 250$ MeV). Right: The solid curve represents  the NLO solution  with $\Lambda= 250$ MeV, while the  long dashed curve the NNLO solution with a value of $\Lambda = 235$ MeV.
}
\label{aperb} 
\label{comparisons}
\end{figure}
\end{center}

In these proceedings, we call perturbative evolution the renormalization group equations (RGE)  that follow from an analysis of the theory as a perturbative expansion in Feynman diagrams with $m$ loops leading to logarithmic corrections of the ratio  'momentum invariant to mass scale', i.e. $\left( \alpha_s \log(P^2/M^2) \right)^m$.
%
At N$^m$LO the scale dependence of the coupling constant is given by

$$\frac{d \, \alpha_s (Q^2)/4\pi}{ d(\ln \;Q^2)}  = \beta_{\mbox{\tiny N}^m\mbox{\tiny LO}}(\alpha_s) =\stackrel{m}{\sum_{k=0}} \left( \frac{\alpha_s}{4\pi}\right)^{k+2} \beta_k\quad.$$
We  show here the solution to $m=1$,  i.e., NLO
\begin{eqnarray}
\beta_0 = \;\; 11\;\; - \;\;\;\frac{2}{3}\, n_f \nonumber \quad ,&& \quad
\beta_1 =  \;102\;\, -\; \;\frac{38}{3} \,n_f \nonumber \quad ,
\end{eqnarray}
where $n_f$ stands for the number of effectively massless quark flavors and $\beta_k$ denote the coefficients of the usual four-dimensional $\overline{MS}$ beta function of QCD.
The evolution equations for the coupling constant can be integrated out exactly leading to
\begin{eqnarray}
 \ln (Q^2/\Lambda_{\mbox{\tiny LO}}^2)&&=\frac{1}{\beta_0 \,\alpha_{\mbox{\tiny LO}}/4\pi } \quad ,\nonumber\\
 \ln (Q^2/\Lambda_{\mbox{\tiny NLO}}^2) &&= \; \frac{1}{\beta_0 \,\alpha_{\mbox{\tiny NLO}}/4\pi} +  \frac{b _1}{\beta_0} \ln\left (\beta_0 \frac{\alpha_{\mbox{\tiny NLO}}}{4\pi}\right)- \frac{b _1}{\beta_0} \ln \left(1 + b_1 \frac{\alpha_{\mbox{\tiny NLO}}}{4\pi}\right) \quad ,
\label{exact}
\end{eqnarray}
%
 %
where $b_k = {\beta_k / \beta_0}$. These equations, except the first, do not admit closed form solution for the coupling constant, and we have solved them numerically. We show their solution, for the same value of $\Lambda = 250$ MeV, in  Fig. \ref{aperb}. 

\begin{center}
\begin{figure}[tb]
\includegraphics[scale=0.66]{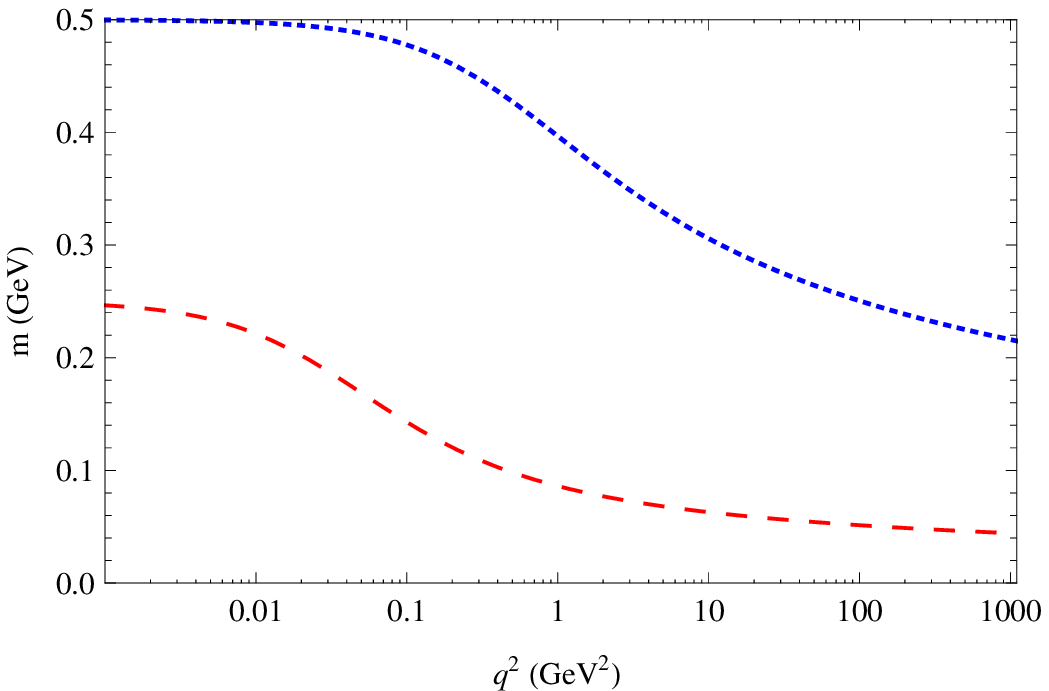}
\includegraphics[scale=0.7]{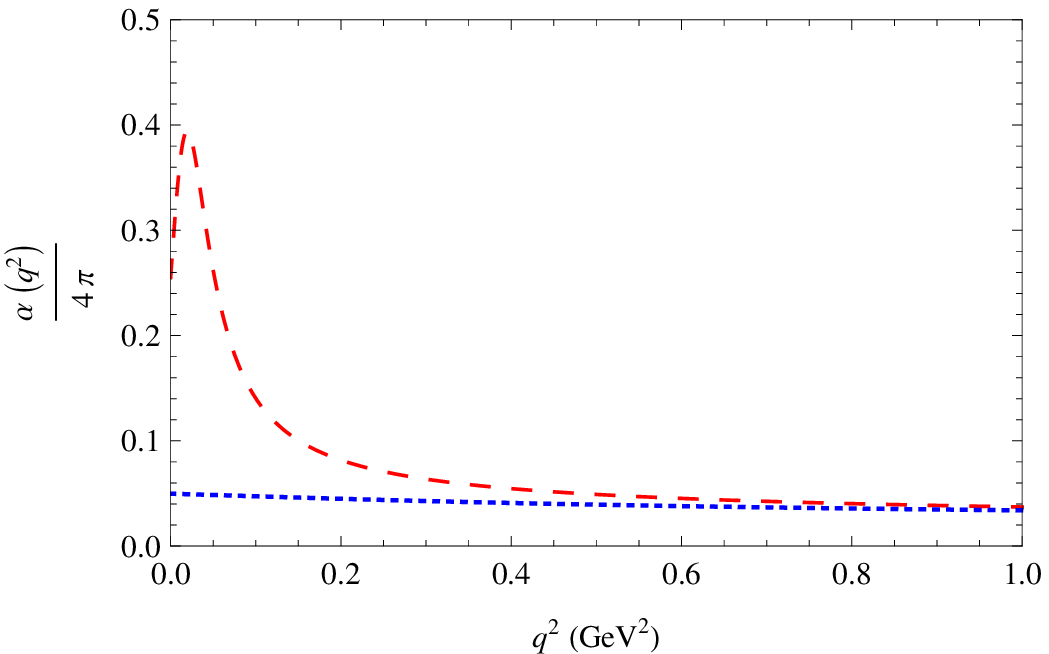}
\caption{ Left: The dynamical gluon mass with a logarithmic running for two sets of parameters, the small mass scenario ($\Lambda =250$ MeV, $m_0 = 250$ MeV, $\rho =1.5$) is shown by the dashed curve; the high mass scenario ($\Lambda =250$ MeV, $m_0 = 500$ MeV, $\rho =2.0$) by the dotted curve. Right: The running of the effective coupling.}
\label{fmass}
\label{alphaJ}
\end{figure}
\end{center}

We see in Fig.~\ref{aperb} (left) that the NLO and NNLO solutions agree quite well even at very low values of $Q^2$ and in  Fig.~\ref{comparisons} (right) that they agree  very well if we change  the value of $\Lambda$ for the NNLO slightly, confirming the fast convergence of the expansion.  This analysis concludes, that even close to the Landau pole, the convergence of the perturbative expansion is quite rapid, specially if we use a different value  of $\Lambda$ to describe the different orders, a feature which comes out from the fitting procedures.
{\it This fast convergence ensures that perturbative evolution can still be used at rather low scales.} However, when entering the non-perturbative regime, other mechanisms take place that influence the QCD evolution. That is what we will call here non-perturbative evolution.
As for the other mechanisms entering into play at low energy, we will consider here, in a first qualitative approach, the consequences of a dynamical gluon mass~ \cite{Cornwall:1981zr,Aguilar:2006gr,Binosi:2009qm,Aguilar:2008xm} on the framework of QCD evolution.  Even though  the  gluon is massless  at the  level  of the  fundamental  QCD Lagrangian, and  remains massless to all order in perturbation theory, the non-perturbative QCD dynamics  generate  an  effective,  momentum-dependent  mass,  without affecting    the   local    $SU(3)_c$   invariance,    which   remains intact.  
At the level of the Schwinger-Dyson equations
the  generation of such a  mass is associated with the existence of {\it infrared finite solutions} for the gluon propagator. Such solutions may  be  fitted  by     ``massive''  propagators  of   the form 
$\Delta^{-1}(Q^2)  =  Q^2  +  m^2(Q^2)$;
$m^2(Q^2)$ can be understood as a constituent gluon mass, and depends non-trivially  on the momentum  transfer $Q^2$.
One physically motivated possibility, which we shall use in here, is  the logarithmic mass running, which is defined by
\begin{equation}
m^2 (Q^2)= m^2_0\left[\ln\left(\frac{Q^2 + \rho m_0^2}{\Lambda^2}\right)
\bigg/\ln\left(\frac{\rho m_0^2}{\Lambda^2}\right)\right]^{-1 -\gamma}.
\label{rmass}
\end{equation}
When $Q^2\to 0$ one has $m^2(0)=m^2_0$. Even though in principle we do not have 
any theoretical constraint that would put an upper bound to the value of $m_0$, 
phenomenological estimates place it in the range $m_0 \sim \Lambda - 2 \Lambda$~\cite{Bernard:1981pg,Parisi:1980jy}. The other parameters were fixed at $\rho \sim 1-4$,  ${(\gamma)} = 1/11$ \cite{Cornwall:1981zr,Aguilar:2007ie,Aguilar:2009nf}. The (logarithmic) running of $m^2$, shown in Fig. \ref{fmass} for two sets of parameters, is   associated with the 
the { gauge-invariant non-local} condensate of dimension two obtained through the minimization
of $\int d^4 x ( A_{\mu})^2$ over all gauge transformations. 

Also, the dynamical gluon mass generation leads  to the "well-established" freezing of the QCD running coupling constant.
 In effect, the  non-perturbative  generalization  of $\alpha_s(Q^2)$ comes, here, in the form
\begin{equation}
\frac{\alpha_{\mbox{\tiny NP}}(Q^2)}{4\pi} = \left[\beta_0 \ln \left(\frac{Q^2 +\rho m^2(Q^2)}{\Lambda^2}\right)\right]^{-1} ,
\label{alphalog}
\end{equation}
where NP stands for Non-Perturbative. The zero gluon mass limit leads to the LO perturbative coupling constant momentum dependence.
The $m^2(Q^2)$ in the argument of the logarithm 
tames  the   Landau pole, and $\alpha_s(Q^2)$ freezes 
at a  finite value in the IR, namely  
\mbox{$\alpha_s^{-1}(0)/4\pi= \beta_0 \ln (\rho m^2(0)/\Lambda^2)$} \cite{Cornwall:1981zr,Aguilar:2006gr,Binosi:2009qm} as can be seen in Fig. \ref{alphaJ} for the same two sets of parameters.
%
We notice the numerical similarity between the perturbative and non-perturbative approaches.  As shown in Fig.~\ref{alpha}, the coupling constant in the perturbative and non-perturbative approaches  are close in size for reasonable values of the parameters from very low $Q^2$ onward ( $Q^2 > 0.1$ GeV$^2$). It shows that, despite the vicinity of the Landau pole to the hadronic scale, the perturbative expansion is quite convergent and agrees with the non-perturbative results for a wide range of parameters.

\section{Non-perturbative Evolution and the Hadronic Scale}

The strong coupling constant plays a central role in the evolution of parton densities.
Let us see how to understand the hadronic scale in the language of models of hadron structure.  We  use, to clarify the discussion, the original bag model, in its most naive description, consisting of a cavity of perturbative vacuum surrounded by non-perturbative vacuum. The non-perturbative vacuum is endowed with a pressure which keeps the cavity size finite. The quarks and gluons are modes in the cavity satisfying certain boundary conditions, for arbitrary product $mR$ (with  $m$ the quark mass and $R$ the radius). The most simple scenario consists only of valence quarks (antiquarks) as Fock states building a colorless state describing the hadron. 
For the lowest mode, the equilibrium for a system of 3 massless quarks is found at
$
R={4\times 2.04}/{M_P},
$
leading to a radius $R>1.$ fm, the size that is required to generate a proton mass from massless quarks.  The associated minimum momentum for the system is then $k_{min, syst}=1/2\,R > 100$ MeV. This is the dynamical scale of the model, a non-perturbative scale arising from the confinement mechanism, which ultimately should be determined by $\Lambda_{QCD}$ if the model were to be derived from the theory, i.e. $B$, the bag pressure, should be related to the only scale of the theory $\Lambda_{QCD}$.

\begin{center}
\begin{figure}
\includegraphics[scale= .70]{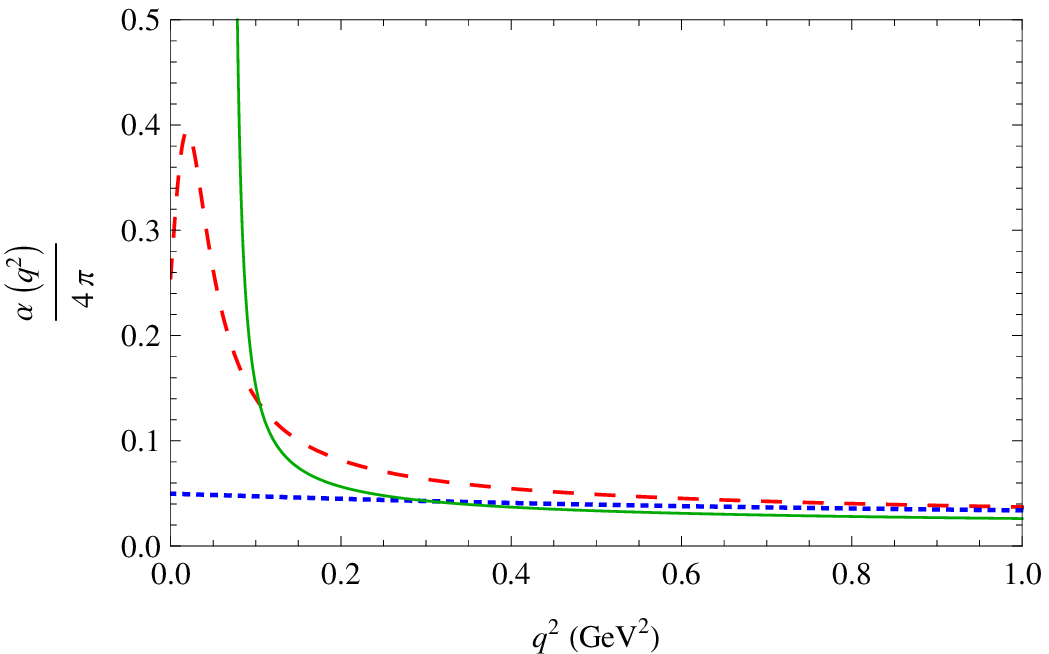}
\includegraphics[scale= .68]{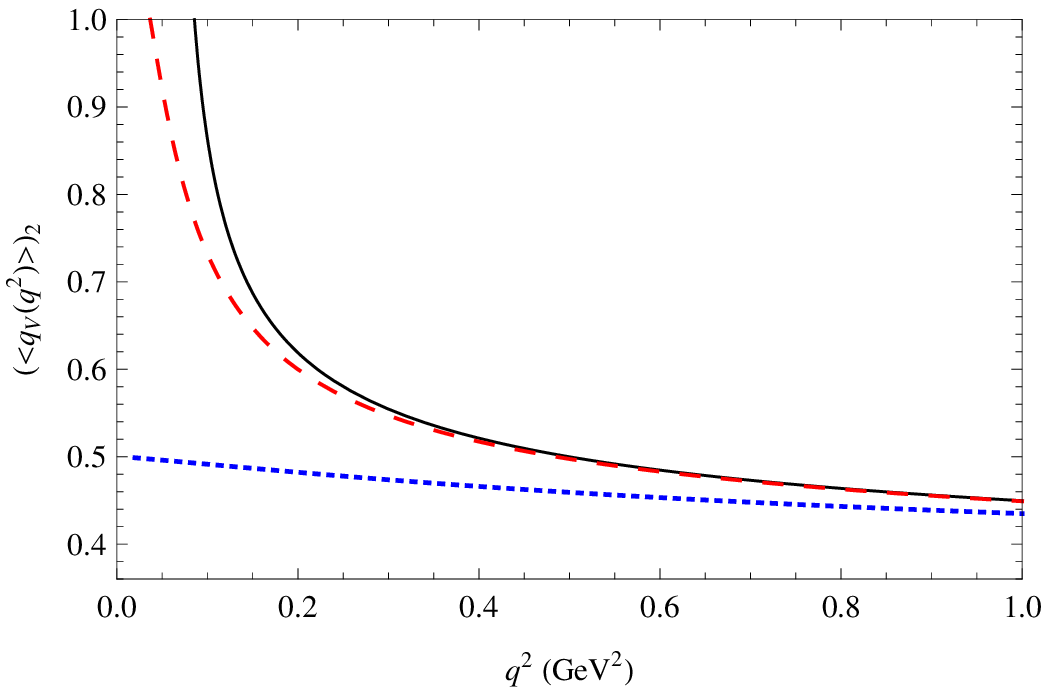}
\caption{ Left: The running of the effective coupling. The dotted and dashed curves represent the non-perturbative evolution with the parameters used above. The solid curve shows the NNLO evolution with $\Lambda = 250$ MeV.
Right: The evolution of the second moment of the valence quark distribution. The solid curve represents the perturbative LO approximation. }
\label{alpha}
\label{qval}
\end{figure}
\end{center}

The bag model is designed to describe fundamentally static properties, but in QCD all matrix elements must have a scale associated to them as a result of the RGE  of the theory.  A fundamental step in the development of the use of hadron models for the description of  properties at high momentum scales was the assertion that all calculations done in a model should have a  RGE scale associated to it \cite{Jaffe:1980ti}. {\it The momentum distribution inside the hadron is only related to the dynamical scale and not to the momentum governing the RGE.} Thus a model calculation only gives a boundary condition for the RG evolution as can be seen for example in the LO evolution equation for the moments of the valence quark distribution 
\begin{equation}
\langle q_v(Q^2)\rangle_n = \langle q_v(\mu_0^2)\rangle_n \left(\frac{\alpha_s{(Q^2)}}{\alpha_s{(\mu_0^2)}}\right)^{d^n_{NS}},
\label{moments}
\end{equation}
where $d^n_{NS}$ are the anomalous dimensions of the Non Singlet distributions.  Inside, the dynamics  described by the model is unaffected by the evolution procedure, and the model provides only the expectation value,  $\langle q_v(\mu_0^2)\rangle_n$, which is associated with the hadronic scale. 
 As mentioned in the first Section, the hadronic scale  is related to the maximum wavelength at which the structure begins to be unveiled.
This explanation goes over to non-perturbative evolution. The non-perturbative solution of the Dyson--Schwinger equations results in the appearence of an infrared cut-off in the form of a gluon mass which determines the finiteness of the coupling constant in the infrared. The crucial statement is that the gluon mass does not affect the dynamics inside the bag, where perturbative physics is operative and therefore our gluons inside will behave as massless. However, this mass will affect the evolution as we have seen in the case of  the coupling constant. The generalization of the coupling constant results to the structure function imply that the LO evolution Eq.~(\ref{moments})  simply changes by incorporating  the non-perturbative coupling constant  evolution Eq.~(\ref{alphalog}). 
The results are shown in Fig.~\ref{qval} (right).

The non-perturbative results, for the same parameters as before, are quite close to those of the perturbative scheme and therefore we are confident that the latter is a very good approximate description.  We note however, that the corresponding hadronic scale, for the sets of parameters chosen, turns out to be slightly smaller than in the perturbative case ($\mu_0^2 \sim 0.1$ GeV$^ 2$), even  for small gluon mass $m_0 \sim 250$ MeV and small $\rho \sim 1$.    One could reach a pure valence scenario at higher $Q^2$ by forcing the parameters but at the price of generating a singularity in the coupling constant in the infrared associated with the specific logarithmic form of the parametrization. We feel that this strong parametrization dependence and the singularity are non physical since the fineteness of the coupling constant in the infrared is a wishful outcome of the non-perturbative analysis. In this sense, the non-perturbative approach  seems to favor a scenario where at the hadronic scale we have not only valence quarks but also gluons and sea quarks \cite{Scopetta:1997wk,Scopetta:1998sg}. In other words, in order to get a scenario with only valence quarks, we are forced to very low gluon masses and very small values $\rho$, while a non trivial scenario allows more freedom in the choice of parameters.
%
%


\section{Non-perturbative Evolution and Final State Interactions.}

The TMDs are the set of functions that depend on both the Bjorken variable $x$, the intrinsic transverse momentum of the quark $|\vec{k}_{\perp}|$
as well as on the scale $Q^2$. Just like PDFs, TMDs are fixed by the possible scalar structures allowed by hermiticity, parity and time-reversal invariance. The existence of leading twist final state interactions allows for time-reversal odd functions. Thus by relaxing time-reversal invariance, one defines two additional functions, the Sivers and the Boer-Mulders functions. 
The growing interest for TMDs called for developments of QCD evolution and its application to PDFs, what  has been recently addressed in Ref.~\cite{Aybat:2011ge}. 
In earlier evaluations of T-odd TMDs, the collinear~\footnote{Collinear, in opposition to transverse, refers to schemes where only longitudinal momenta are relevant. Here: PDFs that depend only on Bjorken-$x$ besides their scale dependence.} perturbative evolution formalism has been naively applied to describe the behavior of the T-odd~Transverse~Momentum~Dependent parton~distribution~functions.

In the standard models' approach towards an evaluation of the T-odd distribution functions, T-odd is allowed thanks to the the final state interactions, which are mimicked  by a one-gluon-exchange. This gluon exchange is usually described through the inclusion of a perturbative gluon propagator. It is precisely due to this mechanism that these functions have an explicit dependence in the coupling constant and therefore they are ideal to analyze the physical impact of our discussion. Since perturbative QCD governs the dynamics inside the confining region, there is no need to include a non-perturbative massive gluon  in the form given by Eq.~(\ref{rmass}), inside the bag.  The main effect of the non-perturbative approach here consists in a change of the hadronic scale $\mu_0^2$ and the value of the running coupling constant at that scale, as  clearly illustrated in Fig.~\ref{alpha}. This leads to a rescaling of the Sivers and Boer-Mulders functions through a change of $\alpha_s (\mu_0^2)$.

\begin{center}
\begin{figure}
\includegraphics[scale= 0.6]{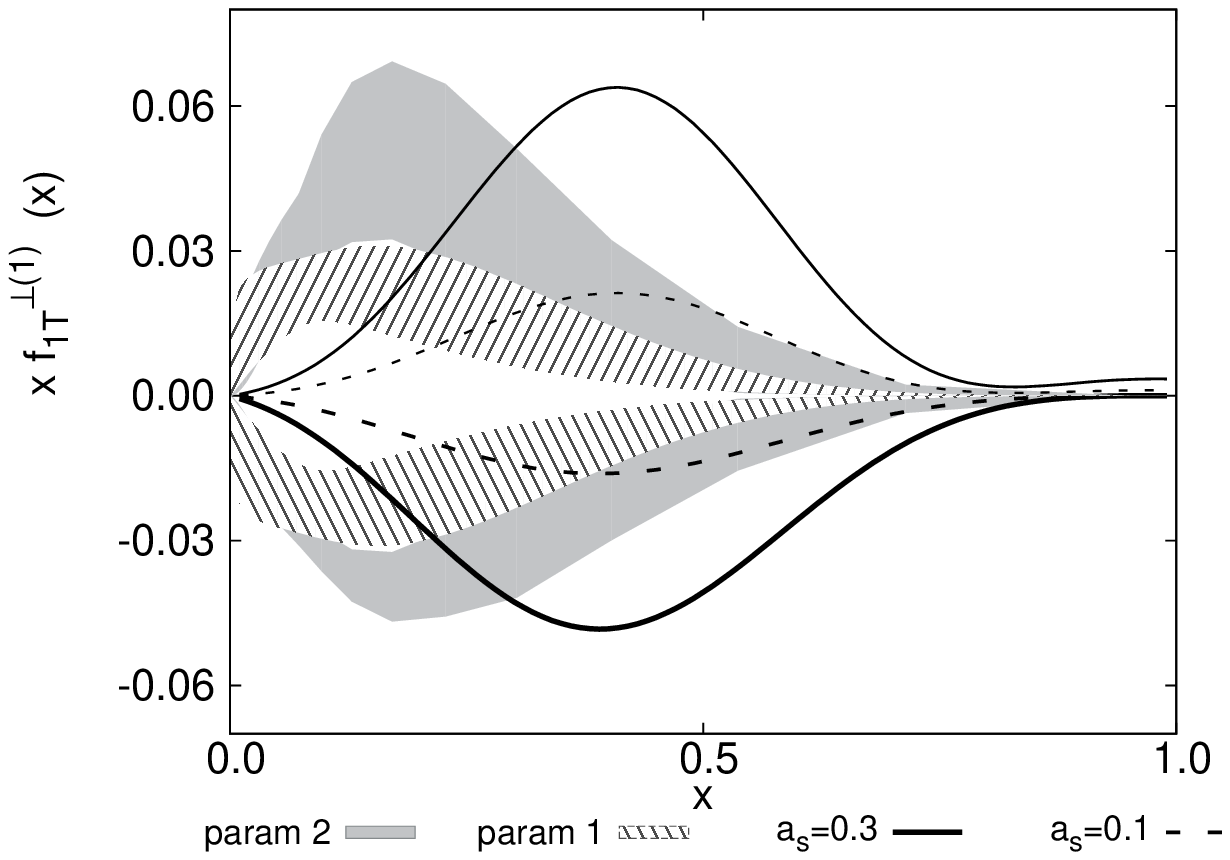}
\includegraphics[scale= 0.6]{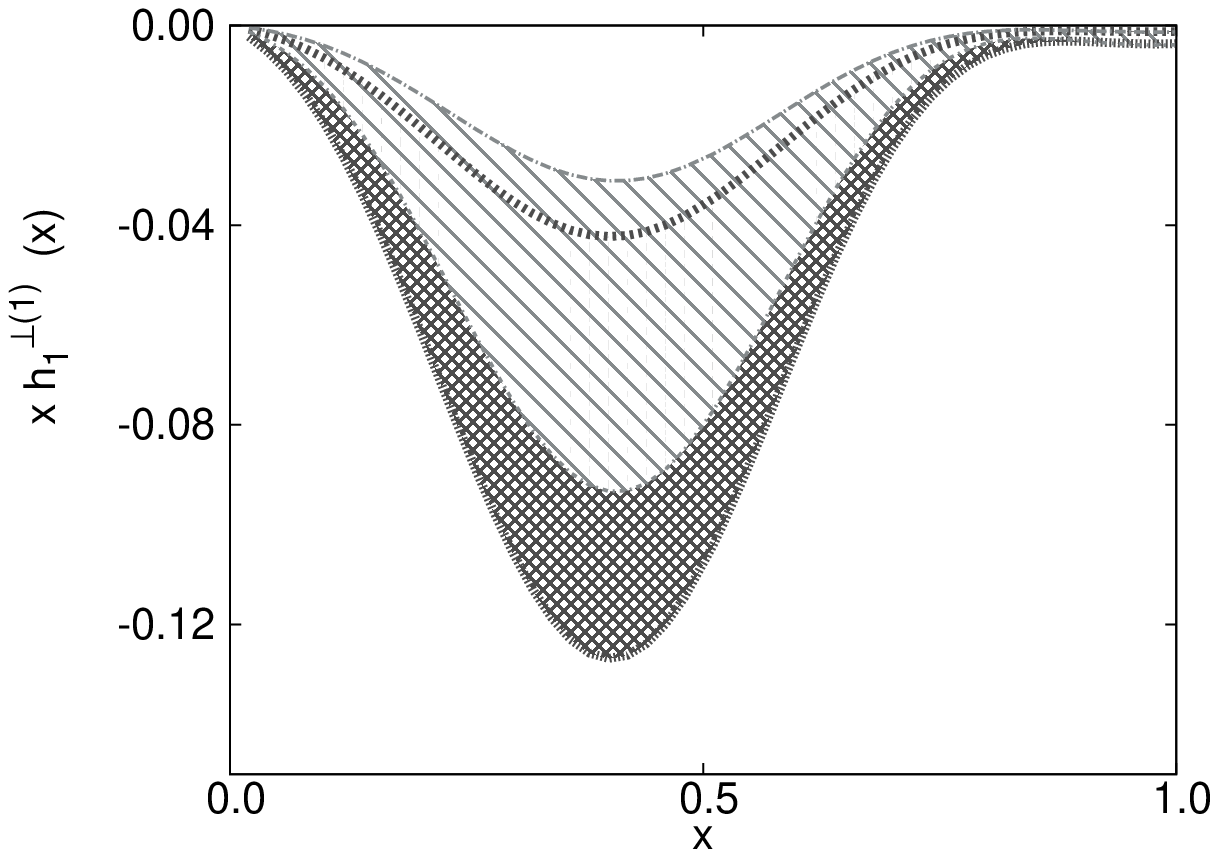}
\caption{The first moment of the Sivers (left) and the Boer-Mulders (right) functions. The results are given for both the $u$ (thick) and $d$ (normal) distributions. The solid (dashed) curves represent the calculation for $a=0.3 (0.1)$. For the plot of the l.h.s., the bands represent the error band for, respectively, the extraction of the Bochum group (full)  and Torino group (stripes).
}
\label{sivers}
\label{bm}
\end{figure}%
\end{center}

In previous calculations~\cite{Courtoy:2008dn,Courtoy:2009pc}, use has been made of the NLO perturbative evolution, with
 ${\alpha_s(\mu_0^2)}/{4\pi}~\sim~0.1$.
Although a solution with this small ${\alpha_s}/{4\pi} $ can be found, with our choice of parameters,  we see, from Fig.~\ref{alpha}, that the coupling constant at the hadronic scale in the non-perturbative  approach and in the NNLO evolution  is consistently larger and lies in the interval
$0.1 <{\alpha_s(\mu_0^2)}/{4\pi} < 0.3$.
Taking into account this range we show the first moments of the Sivers function in  Fig.~\ref{sivers} (left), where we have two extractions from the data at the SIDIS scale~\cite{Collins:2005ie,Anselmino:2008sga}. In order to be able to compare our results to  phenomenology, one should apply the QCD evolution equations. Additionally to the TMD evolution~\cite{Aybat:2011ge}, which should be taken care of properly from now on, one should consider modifications of the evolution equations at low energy -- {\it terra incognita}.
 
If we apply the same band of values of the coupling constant at the hadronic scale to calculation of the Boer-Mulders function we find the results of Fig. \ref{bm} (right).  We see thus how the naive scenario may serve to predict new observables and determine their experimental feasibility. 

The T-odd TMDs have been evaluated in a few models. In most of the models found in the literature  final state interactions are approximated by taking 
into account only the leading contribution due to the one-gluon exchange mechanism. Few non-perturbative evaluations of the T-odd functions have been proposed so far. It is worth noticing that the implementation of the final state interactions is model dependent. The discussion we have presented in this paper is not applicable in general to every scheme. The implementation of the non-perturbative evolution as discussed here might be more complex in other (fully non-perturbative) schemes. So is the description of the confinement mechanism.
%
%
%

\section{Conclusions: Can we extract $\alpha_s$ in the infrared regime from hadronic phenomenology?}

The careful analysis of the previous sections shows that the hadronic scale is close to the infrared divergence (Landau pole) of the coupling constant for conventional values of $\Lambda$.  However, even in this vicinity, the convergence of the N$^m$LO series is very good for the same $\Lambda$ and small modifications of it provide an extremely precise agreement  for all values of $Q^2$ to the right of the pole. Moreover,  an exciting result is that the values obtained by perturbative QCD with reasonable parameters as defined by DIS data, agree with the non-perturbative evaluation of the coupling constant, which is infrared finite, for low values of the gluon mass $m_0 \sim 250$ MeV, and $\rho \sim 1$. 

The naive model for hadron structure that  we have used enables us to control the physics of the problem from the model side as well as to infer from the evolution scenarios that, as expected, the naive pure valence quark scenario is not favored. However, it also shows that the naive scenario may well serve to make predictions, within a reasonably small band, which should not be far from experimental expectations.  Thus, unlike the well-settled formalism used in model calculations, our qualitative approach points out the uncertainty caused by the simple value of the running coupling constant, not to mention the QCD evolution effects.  One could summarize a long time findings, see e.g.~\cite{Traini:1997jz}, by saying that precision in determinations requires sophisticated models, while order of magnitude scenarios can be achieved with naive models. This has a corollary, precise determinations are very model dependent.

In these proceedings we have shown that the hadronic scale can be interpreted not only from the point of view of perturbative evolution, but also from that of non-perturbative momentum dependence of the coupling constant. 
However, both the recent developments in TMD evolution and the still unapprehended non-perturbative evolution formalism should consolidate  a scheme for the matching of moderate energy models to the RGE of the theory.
This scheme should combine an explanation of why the evolution from a low hadronic scale, even in the neighborhood of the Landau pole,  is consistent and can be trusted --- i.e., a non-perturbative approach --- and the behavior at larger $Q^2$ of the defined objects --- i.e., perturbative QCD framework.

The scale fixing procedure uses experimental data and relies on the knowledge of $\alpha_s$ in the sense of perturbative QCD. Vice versa, the new procedure proposed here broadens the ways of analyzing the freezing of the running coupling constant: T-odd TMDs are possible candidates to study the behavior of $\alpha_s$ at intermediate and low $Q^2$, likewise proposed in Ref.~\cite{Deur:2005cf} and Ref.~\cite{Liuti:2011rw}.  A combined analysis of the extractions of the running coupling constant in the infrared region will lead to a novel definition of the effective charge~\cite{future}, following the example of Ref.~\cite{Deur:2005cf}  where the effective coupling constants are phenomenologically inferred from different processes and to calculations based on Schwinger--Dyson equations.

\acknowledgments

I am grateful to V. Vento  for fruitful collaborations and comments. I would like to thank S. Liuti for illuminating discussions. 


\bibliography{courtoy_TNT}


\end{document}